\documentclass{article}
\usepackage{icrctc07}

\title{The Northern Site of the Pierre Auger Observatory}
\shorttitle{The Northern Site}

\authors{%
D. Nitz$^{1}$ for the Pierre Auger Collaboration$^{2}$ \\
}
\shortauthors{The Pierre Auger Collaboration}
\afiliations{$^1$Michigan Technological University, Houghton, MI 49931, USA\\
$^2$Observatorio Pierre Auger, Av. San Mart\'{i}n Norte 304, (5613) Malarg\"{u}e, Argentina}
\email{dfnitz@mtu.edu}

\abstract{The Pierre Auger Observatory is a multi-national project for research
on ultra-high energy cosmic rays. The Southern Auger Observatory in
Mendoza province, Argentina, is approaching completion in 2007 with an
instrumented area of 3,000 $\mathrm{km}^2$. It will accurately
measure the spectrum and composition of ultra-high energy cosmic rays
up to and beyond the predicted GZK feature. We are obtaining
results on the energy spectrum, mass composition and distribution of
arrival directions on the southern sky.  The Northern Auger
Observatory is designed to complete and extend the investigations
begun in the South. It will establish charged particle astronomy and
thus open a new window into the universe. The distribution of arrival
directions of the highest energy events will point the way to
unveiling the almost century old mystery of the origin and nature of
ultra-high energy cosmic rays. Achieving this goal requires collecting
many more events in spite of the steeply falling energy spectrum. The
planned northern site will have an instrumented area of 4,000 square
miles (10,370 $\mathrm{km}^2$) in Southeast Colorado, USA. The
presentation covers the science of charged particle astronomy, the layout and
the technical implementation of the Northern Auger Observatory.}

\begin{document}
\maketitle
%Begin the section.

\section{Introduction}

This paper describes the current design of the Northern Auger
Observatory in the context of the science of ultra-high energy (UHE) cosmic
rays.  The design takes into consideration both the initial science
results from the Southern Auger Observatory and our experience with
the technologies and methods used.

The need for two observatories, one in each hemisphere, for complete
sky coverage at the highest energies was clear from the inception of
the Auger Project.  The Southern observatory site will be 
completed in 2007\cite{icrc299}.

The Southern Observatory with its 1.5 km triangular spacing and an
area of 3,000 $ \mathrm{km}^2 $ will be able to measure accurately the spectrum
and composition from below $10^{18}$ eV to about $10^{20}$
eV\cite{icrc318,icrc594,icrc596,icrc602}. The
statistics above $10^{19}$ eV are sufficient to identify the GZK
feature\cite{greisen,zk}, but marginal for definitive studies of the source
distribution by looking for strong anisotropies in the distribution of
arrival directions\cite{icrc074,icrc075,icrc076}.  However, the data
indicate that the bending power of extragalactic magnetic fields is
small enough to do charged particle astronomy above $10^{19}$ eV and
to therefore be able to observe the sources of ultra-high energy
cosmic rays, given sufficient aperture.  This is the main goal of the
planned Northern Auger Observatory.

 \begin{figure}[tbp]
      \centering
     \includegraphics[width=0.90\columnwidth]{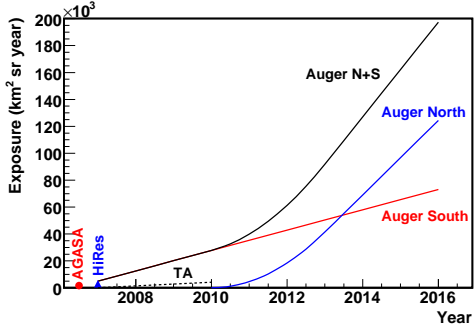}
     \caption{Exposures above $10^{19}$ eV of Auger North and Auger
     South as a function of time.  Also indicated are the expected
     exposures of the Telescope Array\protect{\cite{TA}} and the final
     exposures of the HiRes (monocular)\protect{\cite{hires}} and
     AGASA experiments\protect{\cite{AGASA}}.}
     \label{fig:exposure-time}
 \end{figure}

Auger North will retain the basic functionality and features of Auger
South.  This is important for seamless data integration, e.g. for an
anisotropy analysis on the whole sky.

The Northern hemisphere is chosen to be at roughly the same latitude
and elevation as the Southern site. An important site feature is the
usable area both for initial deployment and possible future expansion.
The chosen site in Southeast Colorado has an initial area of 4,000
square miles (10,370 $\mathrm{km}^2 $),
3.3 times larger than Auger South.

Deployment of the \emph{Surface Detectors}
(SD) is greatly facilitated when they are placed at the corners of
a \emph{square-mile grid}, corresponding to the grid of roads that
exists in Southeast Colorado.
     
Fluorescence Detectors (FD) will again be used for calibration of the
SD, as well as hybrid analysis with accurate composition information
and superior angular resolution on a subset of
events.

\begin{figure}[tbp]
      \centering
   \includegraphics[width=0.90\columnwidth]{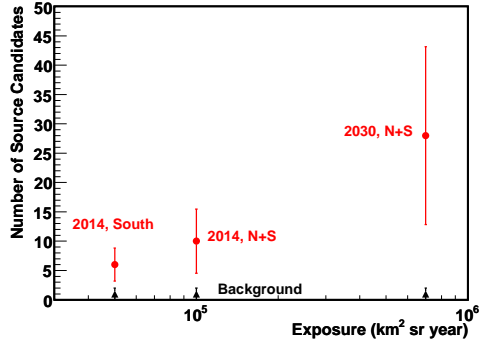}
      \caption{Average number of 5$\sigma$ source candidates over the lifetime of the full Auger Observatory for events
above $10^{20}$ eV.   Black triangles show fake sources from statistical
fluctuations, whereas red circular points show the expected number of source
candidates.}
     \label{fig:Nsources}
 \end{figure}

\section{Science}

The spectrum and composition of UHECRs below $10^{19}$ eV is
most likely the same in both hemispheres as extragalactic particles
below this energy can reach Earth from the entire universe and
galactic ones are isotropized by magnetic fields.

Spectral and composition differences may occur once isotropy is
broken.  As data accumulate above $10^{19}$ eV, departure from
isotropy is expected both from the limited horizon in particle
propagation and the weakening of the effects of cosmic magnetic
fields.  Fig.~\ref{fig:exposure-time} shows the expected accumulated
exposure above $10^{19}$ eV of Auger South, Auger North, and Auger South+North as a
function of time, assuming the construction of Auger North begins in
2009 and is completed in 2012.

Fig. \ref{fig:Nsources} shows the expected number of candidate
``point'' sources detected for Auger South alone by 2014, for both
Auger North and South by 2014, and for Auger North and South combined
by 2030. The number of source candidates was found by generating maps
for each exposure for energies above $10^{20}$ eV and for source
densities of $10^{-5}$ Mpc$^{-3}$, $10^{-4}$ Mpc$^{-3}$, and $10^{-3}$
Mpc$^{-3}$.  The average intensity of each source is adjusted to match
the observed spectrum of cosmic rays.  Isotropic maps were used to
estimate the number of fake sources.  The large exposure and full sky
coverage provided by Auger North will reward us with the detection of
15 to 40 sources by 2030.

In recent years, the great potential for discoveries in UHE neutrino
detections has triggered several experiments, which cover energies
from $10^{14}$ eV up to $10^{26}$ eV.  Given the expected shape of the
cosmogenic neutrino flux, which peaks around $10^{18}$ eV, the
combination of both Auger sites provides the best chance to detect
cosmogenic neutrinos\cite{icrc607}.

\section{Implementation}

\begin{figure}[tbp]
      \centering
     \includegraphics[width=0.55\columnwidth]{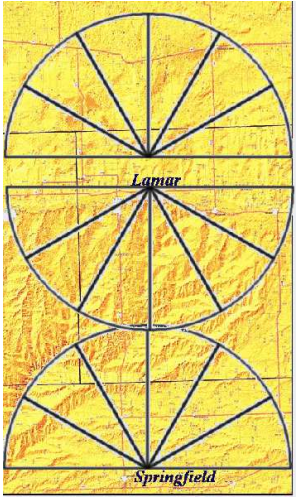}
     \caption{Topographic map of the Auger North site with the fields of view of the 3 fluorescence detector eyes indicated.}
     \label{fig:fd_layout}
 \end{figure}

The layout of the planned Auger North Observatory is indicated in
Fig.~\ref{fig:fd_layout}. Surface detectors are situated on a square-mile
grid covering a 84x48 mile area in the Southeast corner of Colorado.
Three FD eyes overlook the area to provide hybrid coverage.

The square-mile grid layout of the Surface Detector will slightly
decrease the acceptance for small hadron showers yielding an increase
of the threshold energy. The efficiency is $>90\%$ for hadron showers with
5 triggered detectors for energies above $10^{19}$~eV,
while in Auger South it is $3\times10^{18}$~eV.

\subsection{Surface Detectors}

The surface detector electronics planned for Auger North is a natural
evolution of that which is used in Auger South.  The philosophy of
real-time station control software, triggering based upon processing
flash ADC traces, and GPS based time-stamping, which work so well in
Auger South, will remain the same.  However, parts obsolescence
requires a redesign of the electronics.

One of the improvements to the electronics is increased dynamic
range.  Conversely, as
a cost saving measure, the number of PMTs per tank is reduced from
three to one.  Studies in Auger South indicate that this does not
significantly degrade either the triggering or the reconstruction of
the highest energy events.

The integration of the electronics will be increased in
order to reduce cabling and improve reliability.  Increasing the FADC
sampling rate from 40MHz to 100MHz compensates for
the reduction in PMTs. The station controller operating system will
be changed to a variant of real-time Linux.

Unlike the Auger South tanks, Auger North tanks will require thermal
insulation.  One technique being developed is rotationally molded
polyethylene foam insulation on the interior of the tanks.  This
technique is commonly used to increase the stiffness of the walls of
parts being roto-molded.
The Auger North tank design has the main access port in the center for the
single main PMT.

\section{Fluorescence Detector}

The Auger North FD will be split into 3 half eyes, in order to maximize
the number of hybrid events.  The design of the FD eyes is similar to
that of the South.  The HEAT enhancement telescopes\cite{icrc065} serve as a prototype for the North. 

\subsection{Communications Network}

\begin{figure}[tbp]
      \centering
     \includegraphics[width=0.90\columnwidth]{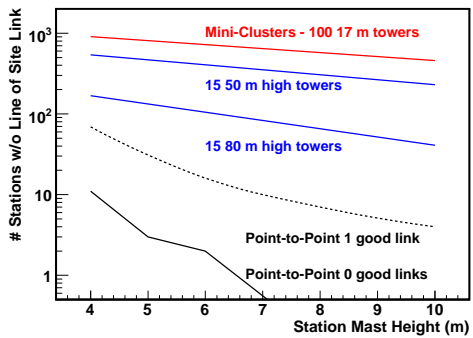}
     \caption{Number of stations (out of 4000) without a line of site communications link.}
     \label{fig:unreachable}
 \end{figure}

Design of the SD communications system for the North takes advantage
of advances during the past decade in wireless network communications.
The southern tanks each communicate independently with local
collectors situated on towers at the FD buildings.  Point-to-point
microwave links to the campus complete the system.  This scheme works
well at the southern site, where the FDs and the towers are situated
substantially higher than the remarkably flat intervening terrain.
The topography of the Southeast Colorado makes this architecture
less suitable for the North.  Fig.~\ref{fig:unreachable} shows the
results of a study, using digital elevation maps (DEM) of
the site, to determine how many of the
4,000 stations would not have a clear line of site to a collector.
Three different scenarios were considered: 1) each tank communicates
to a tower-mounted base station as in Auger South; 2) Mini-clusters,
where each station communicates with a local tower, which are then
networked together; 3) A peer-to-peer network where each station
communicates with one or more of its nearest neighbors. The
peer-to-peer network has many fewer problematic links, and we are thus
pursuing that option for Auger North.

A network of fiber optic cables crossing the site will be used to make
the trunk connections to the central campus facility (instead of the
microwave links used in the South).

\subsection{Data Acquisition}

For Auger South, a comprehensive Central Data Acquisition System
(CDAS) was developed.  CDAS includes both the hardware and software
required to collect incoming data packets from both FD and SD systems,
form and relay triggers, and to save and organize experiment data
online.  Minimal changes will be required to adapt
the existing Auger South CDAS system for Auger North.

\section{Summary}

By pioneering charged particle astronomy, Auger North
will address some of the most compelling questions in science today:
\begin{itemize}
\vspace{-6pt}
\item Where do the highest energy particles that reach the Earth
originate?
\vspace{-6pt}
\item What process in nature can reach such extremely high energies?
\vspace{-6pt}
\item What clues to these particles and their interactions offer about the
universe and its fundamental laws?
\end{itemize}

Answering these questions will transform our view of the most
energetic sites in the present Universe.

\section{Acknowledgments}

The presenter gratefully acknowledges the support of the US Department
of Energy and thanks the Michigan Tech Vice-President for Research for
providing additional funding.

%This is the reference to .bib file (Whitout .bib!)
\bibliography{icrc0180}

\begin{thebibliography}{10}

\bibitem{icrc299}
T.~Suomijarvi [Pierre~Auger Collaboration].
\newblock In {\em Proc. 30th ICRC, ($\#$299)}, 2007.

\bibitem{icrc318}
T.~Yamamoto [Pierre~Auger Collaboration].
\newblock In {\em Proc. 30th ICRC, ($\#$318)}, 2007.

\bibitem{icrc594}
M.~Unger [Pierre~Auger Collaboration].
\newblock In {\em Proc. 30th ICRC, ($\#$594)}, 2007.

\bibitem{icrc596}
M.~Healy [Pierre~Auger Collaboration].
\newblock In {\em Proc. 30th ICRC, ($\#$596)}, 2007.

\bibitem{icrc602}
D.~Barnhill [Pierre~Auger Collaboration].
\newblock In {\em Proc. 30th ICRC, ($\#$602)}, 2007.

\bibitem{greisen}
K.~Greisen.
\newblock {\em Phys. Rev. Lett.}, 16:748, 1966.

\bibitem{zk}
G.T. Zatsepin and V.A. Kuz'min.
\newblock {\em Sov. Phys. JETP Lett.}, 4:78, 1966.

\bibitem{icrc074}
S.~Mollerach [Pierre~Auger Collaboration].
\newblock In {\em Proc. 30th ICRC, ($\#$74)}, 2007.

\bibitem{icrc075}
D.~Harari [Pierre~Auger Collaboration].
\newblock In {\em Proc. 30th ICRC, ($\#$75)}, 2007.

\bibitem{icrc076}
E.~Armengaud [Pierre~Auger Collaboration].
\newblock In {\em Proc. 30th ICRC, ($\#$76)}, 2007.

\bibitem{TA}
H.~Kawai et~al.
\newblock In {\em Proc. 29th ICRC}, pages 8:141--144, 2005.

\bibitem{hires}
R.U.~Abassi et~al.
\newblock {\em ArXiv:astro-ph 0703099v1}, 2007.

\bibitem{AGASA}
M.~Takeda et~al.
\newblock {\em AstroPart. Phys.}, 19:447, 2003.

\bibitem{icrc607}
J.~Alvarez-Muniz [Pierre~Auger Collaboration].
\newblock In {\em Proc. 30th ICRC, ($\#$607)}, 2007.

\bibitem{icrc065}
H.~Klages [Pierre~Auger Collaboration].
\newblock In {\em Proc. 30th ICRC, ($\#$65)}, 2007.

\end{thebibliography}
%This in the bibtex style, is ok.
%\bibliographystyle{plain}
\bibliographystyle{unsrt}

\end{document}